\documentclass[authoryear,11pt,review]{elsarticle} 

\usepackage{mathtools, amsthm, amssymb} 
\usepackage[a4paper, margin=1in]{geometry}
\usepackage{graphicx} 
\usepackage{subcaption}
\usepackage{natbib} 
\usepackage{bm} 
\usepackage{dsfont} 
\usepackage{algorithm} 
\usepackage{algorithmicx} 
\usepackage{algpseudocode} 
\usepackage{array, multirow}
\usepackage{physics}
\usepackage{hyperref}

\begin{document}

\begin{frontmatter}
  \title{A flexible sequential Monte Carlo algorithm for parametric constrained
    regression}

  \tnotetext[t1]{\textcopyright 2019. This preprint is made available under the
    CC-BY-NC-ND 4.0 license.}
  
  \author[sms]{Kenyon Ng\corref{cor1}} \ead{kenyon.ng@research.uwa.edu.au}

  \author[sms,cas]{Berwin A. Turlach}
  \ead{berwin.turlach@uwa.edu.au}

  \author[cas,sph]{Kevin Murray}
  \ead{kevin.murray@uwa.edu.au}

  \address[sms]{Department of Mathematics and Statistics (M019), The
    University of Western Australia, 35~Stirling Highway, Crawley WA 6009}

  \address[cas]{Centre for Applied Statistics (M019), The
    University of Western Australia, 35~Stirling Highway, Crawley WA 6009}

  \address[sph]{School of Population and Global Health (M431), The
    University of Western Australia, 35~Stirling Highway, Crawley WA 6009}

  \cortext[cor1]{Corresponding author}

  \begin{abstract}
    
    An algorithm is proposed that enables the imposition of shape constraints on
    regression curves, without requiring the constraints to be written as closed-form
    expressions, nor assuming the functional form of the loss function. This algorithm
    is based on Sequential Monte Carlo-Simulated Annealing and only relies on an
    indicator function that assesses whether or not the constraints are fulfilled, thus
    allowing the enforcement of various complex constraints by specifying an appropriate
    indicator function without altering other parts of the algorithm. The algorithm is
    illustrated by fitting rational function and B-spline regression models subject to a
    monotonicity constraint. An implementation of the algorithm using \textsf{R} is
    freely available on GitHub.

  \end{abstract}
\end{frontmatter}

\frontmatter

\section{Introduction}
\label{sec:introduction}

Shape constraints, such as monotonicity or convexity, can be required on a regression
curve to ensure its interpretability due to some external theory. For instance, growth
curve modelling often requires monotonicity constraints to be imposed on the response
curve \citep{marsh80age,thornley99modelling}, or shape constrained models can be used to
estimate dose-response curves \citep{kelly90a}, cumulative distribution functions
\citep{ramsey98estimating} and various response curves in economics that are convex by
some underlying theory \citep{semiparametric91matzkin}. Therefore, fitting a shape-constrained
regression model becomes a constrained optimisation problem, where the parameter space
of the regression coefficients is appropriately restricted.

Despite the importance of restricting the shape of the regression curves, current shape
constrained models are mainly concentrated on polynomial or spline models due to their
relative ease of implementation. For instance, when considering monotonicity in
polynomial models, apart from directly minimising the loss function subject to a
restricted parameter space \citep{bon17fitting}, such constraints can be achieved
through an appropriate model parameterisation \citep{elphinstone83a, murray13revisiting,
  murray16fast, manderson17monotone}. On the other hand, when using penalised splines,
the monotonicity or convexity conditions for models with truncated power bases (up to
quadratic spline) or B-spline bases can be written as a system of linear inequality
constraints \citep{leitenstorfer07generalized, brezger08monotonic,
  hazelton11semiparametric, meyer11bayesian, meyer12constrained, pya15shape}. However,
shape-constrained models other than these are less common, partly due to the difficulty
of deriving appropriate closed-form expressions required for enforcing the
constraints. This motivates us to develop a flexible method that is capable of imposing
shape constraints on a wide variety of regression models, even in the absence of
closed-form expressions of the constraints.

We propose a new constrained optimisation method that works on any constraints that are
presentable as an indicator function. Such constraints are not enforced by exploiting
the mathematical expressions that describe them; rather, the feasibility of an estimate
is only known from the indicator function. Consequently, our method is more versatile
than those that assume specific structures on the constraints. To our knowledge, the
only other optimisation method developed with the black-box constraint property is the
Constrained Estimation Particle Swarm Optimisation \citep[CEPSO,][]{wolters12particle}
algorithm. However, CEPSO requires a pilot estimate to operate, which is usually the
unconstrained global minimum. Since the unconstrained global minimum is often difficult
to obtain, especially when there are multiple local minima in the loss function, there
are some circumstances when CEPSO may be unsuitable. Our method is based on Sequential
Monte Carlo-Simulated Annealing \citep[SMC-SA, ][]{zhou13sequential}, and will overcome
the need for a pilot estimate.

The remainder of the paper is outlined as follows: In Section \ref{sec:problem-spec}, we
will formally define the optimisation problem and discuss the challenges of solving
this. The details of the constraint-augmented SMC-SA are discussed in Section
\ref{sec:optimisation}. In Section \ref{sec:applications}, the performance of our method
will be assessed by fitting regression models on both simulated and real-world
datasets. We conclude and discuss our algorithm in Section~\ref{sec:discussion}.

\section{Problem specification}
\label{sec:problem-spec}
Constrained regression analysis is usually analogous to finding a \textit{state}
$\bm{\theta}^* \in \mathcal{S} \subset \mathbb{R}^d$, where the feasible set
$\mathcal{S}$ is not necessarily finite or convex, such that a \textit{loss function}
${\ell:\mathbb{R}^d \rightarrow \mathbb{R}}$ is minimised
\begin{equation*}
  \label{eq:problem}
  \ell(\bm{\theta}^*) \leq \ell(\bm{\theta}) \quad \text{for all $\bm{\theta} \in \mathcal{S}$.}
\end{equation*}
For the purpose of this study, we assume that $\bm{\theta}^*$ is unique.

The classic approach of solving constrained optimisation problems is to introduce
Lagrange multipliers into the loss function, and solve the system of equations attained
from partially differentiating the transformed function with respect to the parameters
and Lagrange multipliers \citep[see for example][]{nocedal06numerical}. In the usual
case of minimising a complicated loss function, iterative algorithms may be the more
sensible choice \citep[see for example][]{romeijn94simulated, andreani13a,
  reddi15large-scale}.

However, most of the existing constrained optimisation methods are restricted to only
solving problems with particular types of constraints, and are consequently not suitable
for the development of a flexible constrained optimisation methodology. Moreover, there
are constraints which cannot be written down as a finite set of closed-form
expressions. For example, fitting a least squares continuous rational function requires
its denominator polynomial to be non-zero over the region of interest, implying that
there are infinitely many non-equalities to be considered. It is therefore difficult, if
not impossible, to implement any of the traditional methods for this problem.

\section{Optimisation}
\label{sec:optimisation}
It is well known that the Boltzmann distribution of the form
\begin{equation*}
  \label{eq:boltzmann}
  \pi(\bm{\theta}) \propto \exp \left( - \dfrac{\ell(\bm{\theta})}{T} \right)
\end{equation*}
will converge to a degenerate distribution whose mass is concentrated on the global
minimum of~$\ell(\bm{\theta})$ when the temperature $T$ converges to 0. Similarly for
constrained optimisation problems, it can be shown that this will still hold when the
solution space is restricted. That is, the distribution
\begin{equation}
  \label{eq:boltzmann-cons}
  \pi(\bm{\theta}) \propto \exp \left( - \dfrac{\ell(\bm{\theta})}{T} \right)
  \, \mathds{1}_{\mathcal{S}},
\end{equation}
where $\mathds{1}_\mathcal{S}$ is an indicator for $\bm{\theta} \in \mathcal{S}$, will
converge to a degenerate distribution concentrated on the constrained global minimum
$\bm{\theta}^*$ \citep{romeijn94simulated}. In light of these results, a constrained
optimisation problem can essentially be solved by drawing samples
from~\eqref{eq:boltzmann-cons} with~\mbox{$T \rightarrow 0$}, a fact employed by the
algorithm we propose.

\subsection{Simulated annealing}
\label{sec:sa}
The Metropolis algorithm \citep{metropolis53equation}, a Markov chain Monte Carlo (MCMC)
technique, can be employed to draw samples from \eqref{eq:boltzmann-cons}, in a similar
fashion to simulated annealing \citep{kirkpatrick83optimization,
  locatelli00simulated}. More specifically, we sample from a sequence of Boltzmann
distributions
\begin{equation}
  \label{eq:boltzmann_sequence}
  \pi_k(\bm{\theta}) \propto \exp \left( - \dfrac{\ell(\bm{\theta})}{T_k} \right)
  \, \mathds{1}_\mathcal{S},
  \quad k = 1, 2, 3 \ldots
\end{equation}
that starts from a Boltzmann distribution with high temperature and eventually converges
to the degenerate distribution of interest (i.e.~$T_k \rightarrow 0$). Simulated
annealing first simulates a Markov chain with invariant $\pi_1$, then proceeds to
simulate the rest of $\pi_k$ with fresh Markov chains using the sample from the
preceding $\pi_{k-1}$ as starting states. This construction of sampling from a
distribution sequence, rather than direct sampling from \eqref{eq:boltzmann-cons} with
low~$T$, is necessary to facilitate efficient sampling due to the low acceptance
probability of the Markov chain when the temperature is low. This can be observed
evidently from the formula of the acceptance probability for each~$\pi_k$
\begin{equation}
  \label{eq:acceptance}
  p_k = \min{ \left\{ 1, \exp \left( -\dfrac{\ell(\tilde{\bm{\theta}}) -
          \ell(\bm{\theta}_{k-1})}{T_k} \right) \,
      \mathds{1}_\mathcal{S}(\tilde{\bm{\theta}}) \right\} },
\end{equation}
where $\tilde{\bm{\theta}}$ and $\bm{\theta}_{k-1}$ denote proposed and previous states
respectively. At low temperatures, even a small difference of $\ell(\bm{\theta})$ will
lead to the exponential function evaluating to a large negative value.

Since simulated annealing minimises $\ell(\bm{\theta})$ by sampling from
\eqref{eq:boltzmann_sequence}, there are two decisions to be made: the number of samples
drawn from $\pi_k$ before moving to $\pi_{k+1}$, and the choice of temperature $T_k$ at
each iteration. Naturally when sampling from $\pi_k$, the ideal situation will be
simulating its corresponding Markov chain until the invariant distribution of the chain
is achieved, and using the sample from $\pi_k$ as the starting state of the subsequent
Markov chain to sample from $\pi_{k+1}$. This strategy is clearly impractical in general
due to the large amount of transitions required for each Markov chain to achieve
stationarity; rather, the common practice is to perform only a single transition for
each $k$. Consequently, those Markov chains never achieve their stationarity, and some
conditions (such as the choice of $T_k$ and the proposal distribution) have to be
followed to ensure that the whole algorithm will still minimise $\ell(\bm{\theta})$ when
$k \rightarrow \infty$ \citep{locatelli00simulated}. For more detailed analysis and
implementation of constrained simulated annealing with compact $\mathcal{S}$, we refer
to \citet{romeijn94simulated} and \citet{locatelli02simulated}.

While simulated annealing will converge to the global minimum when
$k \rightarrow \infty$, in practice one can only carry out a finite number of
iterations. This may lead the algorithm to converge to a local minimum if it starts from
a poor state and stops prematurely. However, with the abundance of parallel computing
power available nowadays, this issue can be mitigated by performing multiple independent
simulated annealings that start from different states. For the rest of this paper, we
will refer to this technique as multi-start SA, which starts from~$N$ different
locations, leading to the following set of search chains
$\{ \bm{\theta}^j_k \}^K_{k=0}, \, j = 1,2, \ldots, N$ after $K$ iterations. Arguably,
this strategy of starting from multiple states is widely used by practitioners when
multiple local minima are present in the loss function, in the hope that at least one of
the search chains will converge to the global minimum. As we will argue in the
following, multi-start SA can be improved by utilising the information extracted from
$\{ \bm{\theta}^j_{k} \}^N_{j = 1}$ at each iteration, which is the motivation of
SMC-SA.

\subsection{Sequential Monte Carlo-Simulated Annealing}
\label{sec:smc-sa}
An alternative approach to sampling from~\eqref{eq:boltzmann_sequence} is the Sequential
Monte Carlo (SMC) sampler proposed by \citet{del06sequential} and formalised by
\citet{zhou13sequential}, who then called the resulting algorithm Sequential Monte
Carlo-Simulated Annealing (SMC-SA). Intuitively speaking, SMC-SA is a multi-start SA
algorithm combined with the idea of importance sampling. Prior to the start of Markov
transitions at each iteration, the importance sampling step will only select the states
with lower loss function values for further optimisation, and discard the rest to save
computational resources. From a MCMC perspective, all states
$\{ \bm{\theta}^j_{k-1} \}^N_{j = 1}$ are weighted and sampled at the start of each
iteration to resemble $\pi_k$ before performing a Markov transition, such that the
single-transition Markov chain can achieve its invariant distribution (i.e.~$\pi_{k}$)
more easily.

We will discuss SMC-SA as a two-step process, namely importance sampling and the SA-move.
\begin{enumerate}
\item Importance sampling: The idea of the importance sampling step is to filter out
  states with greater loss function values, which are less likely to lead to the global
  minimum. This is implemented by sampling \textit{promising states}
  $\{\bm{\gamma}^j_k\}^{N}_{j=1}$ from the pool of previous
  states~$\{\bm{\theta}^j_{k-1}\}^{N}_{j=1}$ at the start of each iteration, according
  to the weights~$\{w^j_k\}^N_{j=1}$ assigned to each~$\bm{\theta}^j_{k-1}$
  \begin{equation}
    \label{eq:weights}
    w^j_k \propto
    \begin{dcases}
      \exp \left( - \dfrac{\ell(\bm{\theta}^j_0)}{T_1} \right), & k = 1; \\
      \exp \left( -\ell(\bm{\theta}^j_{k-1}) \left( \dfrac{1}{T_k} - \dfrac{1}{T_{k-1}}
        \right)
      \right), & k > 1. \\
    \end{dcases}
  \end{equation}
  The eventual output from the importance sampling step is a set
  $\{\bm{\gamma}^j_k\}^{N}_{j=1}$.

\item SA-move (Markov kernel): The $\{\bm{\gamma}^j_k\}^{N}_{j=1}$ from the importance
  sampling step are then transitioned once with a Markov kernel with invariant
  distribution of $\pi_k$. For each $\bm{\gamma}^j_k$, a candidate state is drawn from a
  symmetrical random walk proposal $q(\bm{\theta} | \bm{\gamma}^j_k)$ which is centred
  on $\bm{\gamma}^j_k$, and performs an accept-reject routine according to
  \eqref{eq:acceptance} to determine~$\bm{\theta}^j_{k}$. This step is essentially
  performing an iteration of the Metropolis algorithm with $\bm{\gamma}^j_k$ as the
  starting state.
\end{enumerate}
The pseudocode of SMC-SA is outlined in Algorithms~\ref{algo:samove} and \ref{algo:smcsa}.

\begin{algorithm}
  \caption{SA-Move (Markov kernel $K_k(\bm{\theta}_{k-1},
    \bm{\theta}_{k})$)} \label{algo:samove}
  \begin{algorithmic}[1]
    \Procedure{SA-Move}{$\bm{\theta}_{k-1}$, $T_{k}$}
    \State $\tilde{\bm{\theta}} \gets$ A sample from $q(\bm{\theta} |
    \bm{\theta}_{k-1})$.
    \State $p_{k} \gets$ Evaluate~\eqref{eq:acceptance}.
    \State $\bm{\theta}_k \gets \tilde{\bm{\theta}}$ with probability $p_k$;
    $\bm{\theta}_k \gets \bm{\theta}_{k-1}$ otherwise.
    \State \Return $\bm{\theta}_{k}$
    \EndProcedure
  \end{algorithmic}
\end{algorithm}

\begin{algorithm}
  \caption{Sequential Monte Carlo-Simulated Annealing} \label{algo:smcsa}
  \begin{algorithmic}[1]
    \Procedure{SMC-SA}{$\{\bm{\theta}^j_0\}^N_{j=1}$}
    \For{$k = 1,2,3 \ldots$}
    \State $T_k \gets T(k)$
    \State $\{w^j_k\}^N_{j = 1} \gets$ Evaluate~\eqref{eq:weights}
    \State $\{\bm{\gamma}^{j}_k\}^{N}_{j = 1} \gets$ Resample from
    $\{\bm{\theta}^j_{k-1}\}^N_{j=1}$, with weights $\{w^j_k\}^N_{j=1}$.
    \For{$j = 1,2,3,\ldots,N$}
    \State $\bm{\theta}^j_{k} \gets$ \textsc{SA-Move}($\bm{\gamma}^j_k$, $T_k$)
    \If{$\ell(\bm{\theta}^j_{k}) < \ell(\bm{\theta}^*)$} $\bm{\theta}^* \gets
    \bm{\theta}^j_{k}$
    \EndIf
    \EndFor
    \EndFor
    \EndProcedure
  \end{algorithmic}
\end{algorithm}

\subsubsection{Proposal distribution}
The selection of $q(\bm{\theta} | \bm{\theta}_{k-1})$ can greatly affect the transition
of the chain, and thus the quality of the samples being drawn. More specifically, a
proposal distribution with significant mass outside $\mathcal{S}$ will result in the
chain rarely moving to a new state, as \eqref{eq:acceptance} will evaluate to 0
when~$\tilde{\bm{\theta}}$ is not feasible. In this case, a random walk proposal with
support $\mathcal{S}$, which we refer to as a truncated random walk for the remainder of
this paper, will be a more ideal candidate than a symmetrical random walk as the
acceptance probability will be strictly positive. Using a proposal other than a
symmetrical random walk also implies that \eqref{eq:acceptance} needs to be adjusted
with a ratio of proposal densities evaluated at the current and proposed states
\citep{hastings70monte}
\begin{equation}
  \label{eq:acceptance-cons-mh}
  p_k = \min \left\{ 1, \dfrac{q(\tilde{\bm{\theta}} |
        \bm{\theta}_{k-1})}{q(\bm{\theta}_{k-1} | \tilde{\bm{\theta}})} 
      \exp \left( -\dfrac{\ell(\tilde{\bm{\theta}}) - \ell(\bm{\theta}_{k-1})}{T_k}
        \right) \,
        \mathds{1}_{\mathcal{S}}(\tilde{\bm{\theta}})  \right\} , 
\end{equation}
to preserve the reversibility of the Markov kernel. However, the calculation of
\eqref{eq:acceptance-cons-mh} is not trivial.

In light of the ratio of proposal distributions in \eqref{eq:acceptance-cons-mh} being
intractable, we suggest ignoring this ratio and using \eqref{eq:acceptance} to calculate
acceptance probabilities. This strategy is equivalent to having a Markov chain with a
symmetrical random walk proposal and transitioning the chain until the acceptance
probability is non-zero. A truncated random walk merely rejects the infeasible states at
the proposal stage rather than at the acceptance-rejection stage. Although omitting the
proposal ratio might result in the Markov chains having invariant distributions other
than \eqref{eq:boltzmann_sequence}, we are not interested in estimating the density of
the exact Boltzmann distributions, but rather the states that correspond to high density
regions. The sequence of Boltzmann distributions is only a vehicle to minimise
$\ell(\bm{\theta})$. Our aim can still be achieved as long as the global minimum is
visited by the algorithm.

Sampling from a truncated random walk, especially when the nature of the constraint is
unknown, is difficult and we usually have to use a rejection sampler. In our algorithm,
we use a truncated Gaussian random walk proposal
\begin{equation}
  \label{eq:proposal-cons}
  \tilde{\bm{\theta}} = \bm{\theta}_{k-1} + \bm{\epsilon}_{w}, \quad p(\bm{\epsilon}_{w})
  \propto \exp \left( - \dfrac{ \bm{\epsilon}_{w}^T
      \bm{\epsilon}_{w}}{2 \sigma^2} \right) \, \mathds{1}_{\mathcal{S}}(\tilde{\bm{\theta}}).
\end{equation}
This proposal can be generated with a rejection sampler, by repeatedly adding a
$d$-dimensional Gaussian noise~$\mathcal{N}_d(\bm{0}, \sigma^2 \bm{I})$
to~$\bm{\theta}_{k-1}$ until a feasible~$\tilde{\bm{\theta}}$ is obtained.

When $\bm{\theta}$ is high-dimensional, $\mathcal{S}$ can be small relative to
$\mathbb{R}^d$; thus, a rejection sampler for~\eqref{eq:proposal-cons} is notoriously
inefficient. In this case, we suggest using a $K$-point operator proposal, as shown in
\citet{liang14simulated}, which updates only $K < d$ components of $\bm{\theta}_{k-1}$
in each iteration. The components to be updated are randomly chosen at the beginning of
the Markov transition, and a~\mbox{$K$-dimensional} Gaussian noise
$\mathcal{N}_K(\bm{0}, \sigma^2\bm{I})$ is repeatedly added to the chosen components
until a feasible~$\tilde{\bm{\theta}}$ is obtained, as in the case where $K = d$.

\subsubsection{Cooling schedule}
\label{sec:cooling-schedule-sa}

A cooling schedule is a function that controls the temperature at each iteration, which
in turn affects the algorithm convergence. It should restrict the cooling rate such that
the invariant distributions of adjacent Markov chains do not differ too much, and the
initial states for each chain in \eqref{eq:boltzmann_sequence} are more likely to fall
within the high density region of their respective invariant distributions. The choice
of cooling schedule is usually dependent on the problem at hand. However, in general,
there are two conditions that need to be met for SMC-SA to converge:~$T_k \rightarrow 0$
and $\left| \dfrac{1}{T_k} - \dfrac{1}{T_{k-1}} \right|$ is monotonically decreasing
\citep{zhou13sequential}. Schedules that satisfy both of these conditions include the
\textit{logarithm schedule}, which is suggested in \citet{zhou13sequential} and has the
form
\begin{equation*}
  \label{eq:log-schedule}
  T(k) = \dfrac{\lvert \ell(\bm{\theta}^*_k) \rvert}{\log{(k + 1)}},
\end{equation*}
where $\bm{\theta}^*_k$ is the best state observed up until the $k^{th}$
iteration. However, as described in~\citet{nourani98a}, the cooling rate of the
logarithm schedule can be too conservative, and an alternative is to use a more
stringent \textit{reciprocal schedule}
\begin{equation}
  \label{eq:recip-schedule}
  T(k) = \dfrac{\lvert \ell(\bm{\theta}^*_k) \rvert}{1 + \alpha (k-1)^2},
  \quad 0 < \alpha < 1.
\end{equation}
In our experience, this schedule outperforms the logarithm schedule when using our
modified SMC-SA algorithm in terms of the best minimum achieved for a given number of
iterations, despite $\left| \dfrac{1}{T_k} - \dfrac{1}{T_{k-1}} \right|$ not being
monotonically decreasing.

\subsubsection{SMC-SA and SMC sampler}
The SMC-SA algorithm can also be seen as a special case of the SMC sampler
\citep{del06sequential} that simulates the distribution sequence
in~\eqref{eq:boltzmann_sequence} with the following specifications:
\begin{itemize}
\item the initial distribution is $\pi_0(\bm{\theta}) \propto 1$; 
\item Metropolis MCMC forward kernels $K_k(\bm{\theta}_{k-1}, \bm{\theta}_k)$ are used
  with invariant distributions given by~\eqref{eq:boltzmann_sequence};
\item the backward kernels are
  \begin{equation*}
    L_{k-1}(\bm{\theta}_{k}, \bm{\theta}_{k-1}) = \dfrac{\pi_k(\bm{\theta}_{k-1})
      K_k(\bm{\theta}_{k-1}, \bm{\theta}_k) }{\pi_k(\bm{\theta}_k)};
  \end{equation*}
\item states are resampled after each iteration to avoid degeneracy.
\end{itemize}
In this case, \eqref{eq:weights} gives the weights for each sample in the SMC
sampler. The implementation of~$K_k(\bm{\theta}_{k-1}, \bm{\theta}_k)$ is outlined in
Algorithm~\ref{algo:samove}.

\subsection{Stochastic approximation annealing}
Stochastic approximation annealing (SAA) is another improvement on the simulated
annealing proposed by \citet{liang14simulated}. Rather than sampling from
\eqref{eq:boltzmann_sequence} directly, SAA partitions the parameter space into multiple
regions $L_1, L_2, \ldots, L_m$ according to their loss function values
\begin{gather*}
  L_1 = \{ \bm{\theta} : \ell(\bm{\theta}) \leq l_1 \}, \,
  L_2 = \{ \bm{\theta} : l_1 < \ell(\bm{\theta}) \leq l_2 \}, \ldots, \\
  L_{m-1} = \{ \bm{\theta} : l_{m-2} < \ell(\bm{\theta}) \leq l_{m-1} \}, \,
  L_m = \{ \bm{\theta} : \ell(\bm{\theta}) > l_{m-1} \}
\end{gather*}
where $l_1 < l_2 < \ldots < l_{m-1}$ are arbitrary but fixed. SAA seeks to sample from
the following sequence of partitioned Boltzmann distributions
\begin{equation}
  \label{eq:partitioned_boltzmann}
  \pi_k(\bm{\theta}) \propto \sum^m_{i = 1} \exp \left\{
      -\dfrac{\ell(\bm{\theta})}{T_k} - \zeta^i_k \right\} \, \mathds{1}_{\mathcal{S} \cap
  L_i} , \quad k = 1, 2, 3 \ldots
\end{equation}
which resembles a mixture of Boltzmann distributions truncated to
$L_1, L_2, \ldots, L_m$. The extra hyperparameters $\zeta^i_k$, which vary at each
iteration, control the mixing probability and are updated to encourage the exploration
of partitions that are previously unvisited. \citet{liang14simulated} proved that such
augmentation on the Boltzmann distributions allows SAA to converge using a square-root
schedule as opposed to the logarithm schedule used in standard simulated annealing, thus
achieving a faster convergence rate.

In principle, we can further improve SAA by using an SMC sampler, instead of a
Metropolis-Hastings sampler as suggested in the original implementation in
\citet{liang14simulated}, to simulate the sequence \eqref{eq:partitioned_boltzmann}. We
conjuncture that the resulting algorithm will resemble the following structure in each
iteration:
\begin{enumerate}
\item Perform importance sampling with a new set of weights derived for
  \eqref{eq:partitioned_boltzmann};
\item Perform Markov transition (SA-move) on each state with
  \eqref{eq:partitioned_boltzmann} as the invariant distributions;
\item Update $\zeta$, as in the $\theta$-updating step in \citet{liang14simulated}.
\end{enumerate}

While SAA incorporated with a SMC sampler may converge faster, the convergence
performance of SAA depends on the partition $l_1, \ldots, \l_{m-1}$, which can be
challenging to choose. In light of this issue, a detailed study of incorporating SAA
into our algorithm is left for future research.

\section{Applications}
\label{sec:applications}
In this section, we will discuss the convergence performance of our augmented SMC-SA in
a regression context. We demonstrate our algorithm by fitting rational function models
and B-spline models, both of which are subject to monotonicity constraints. We use
either a least squares estimator or Tukey's biweight estimator, depending on the
dataset. The results are then compared to that from multi-start SA and CEPSO, which will
also serve as our benchmark. This simulation study was conducted in~\textsf{R}
\citep{team18r} on an Intel Core i7-6700 (3.4GHz) machine with 8 gigabytes of RAM. The
code for implementing the algorithm and simulation study is available on
\url{https://github.com/weiyaw/blackbox}.

All of the algorithms which we tested require a set of starting states
$\{ \bm{\theta}^j_0 \}^{1000}_{j = 1}$, rather than a single point in the parameter
space, to operate efficiently. We ran all the algorithms 40 times, each of which started
with different sets of 1000 states. The same set of starting states were used across
different algorithms to ensure a fair comparison. The set of starting states are
generated from the following procedure:
\begin{enumerate}
\item Obtain a crude estimate $\bm{\eta}_0$; \label{prod:step1}
\item Draw a sample from $\text{Cauchy}(\bm{\eta}_0, 2)$ until that sample is
  feasible; \label{prod:step2}
\item Repeat Step \ref{prod:step1} and \ref{prod:step2} for 1000 times.
\end{enumerate}
A Cauchy distribution is preferred due to its heavy tails \citep{meyer03evolutionary},
which allows the set of starting states to cover more areas in $\mathcal{S}$, although
a uniform distribution is more suitable when~$\mathcal{S}$ is bounded.

We tried three different cooling schedules for SMC-SA: the logarithm schedule, and the
reciprocal schedule with $\alpha$ set to 0.85 and 0.95. For multi-start SA, we only used
the logarithm schedule since the reciprocal schedule is too aggressive. The variance
parameter $\sigma^2$ in \eqref{eq:proposal-cons}, the number of states $N$ and the
number of iterations performed were then set according to the cooling schedule used in
the algorithm and are given in Table~\ref{tab:hyperpar}. The $\sigma^2$ was decreased
after each iteration to allow a smaller step size and thus a finer improvement at the
latter stages of the algorithm. We also used a 2-point operator proposal for all the
examples, unless specified otherwise. The number of iterations was chosen to ensure that
the computational efforts across different algorithms were roughly the same. In general,
a more stringent cooling schedule requires less iterations but more samples to converge.

\begin{table}
  \centering
  \caption{The hyperparameters for SMC-SA and multi-start SA, according to the cooling
    schedule used. The $\sigma^2$ was decreased $0.2\%$ and $3\%$ after each iteration
    for the logarithm and reciprocal schedules respectively. The number of states $N$
    and iterations are chosen to ensure each algorithm receives roughly the same
    computational resources. The starting states are duplicated 3 times when
    $N = 3000$.}
  \medskip
  \begin{tabular}{*{1}{l}*{3}{c}}
    \firsthline
    Schedule & $\sigma^2$ & $N$ & Iterations \\
    \hline
    Logarithm & $1 \times 0.998^k$ & 1000 & 3000 \\
    Reciprocal $(\alpha = 0.85/0.95)$ & $1 \times 0.97^k$ & 3000 & 1000 \\
    \lasthline
  \end{tabular}
  \label{tab:hyperpar}
\end{table}

For CEPSO, the algorithm used 2000 iterations and the neighbourhood size for the local
best estimation was set to 200. The methods used to obtain the pilot estimates are
problem-specific and will be discussed later.

\subsection{Rational function models}
\label{sec:rational}
Rational functions are fractions whose numerator and denominator are both
polynomials. They are usually more flexible than polynomials in the sense that they can
describe a given curve with fewer parameters, yet retaining the same degree of accuracy
\citep{newman64rational,ralston01first}. Apart from their superior flexibility, some
physical phenomena can be precisely modelled as a rational function, such as the
Michaelis-Menten kinetics model in biochemistry \citep{piegorsch05analyzing}. However,
rational function models are not continuous at the roots of their denominator, making it
challenging to reliably fit them to data in practical applications; see
Figure~\ref{fig:ht1}.

\subsubsection{A simulated dataset from a hyperbolic tangent function}
\label{sec:hyper-tangent}
\begin{figure}
  \centering
  \begin{subfigure}[t]{0.49\textwidth}
    \centering
    \includegraphics[width=\textwidth]{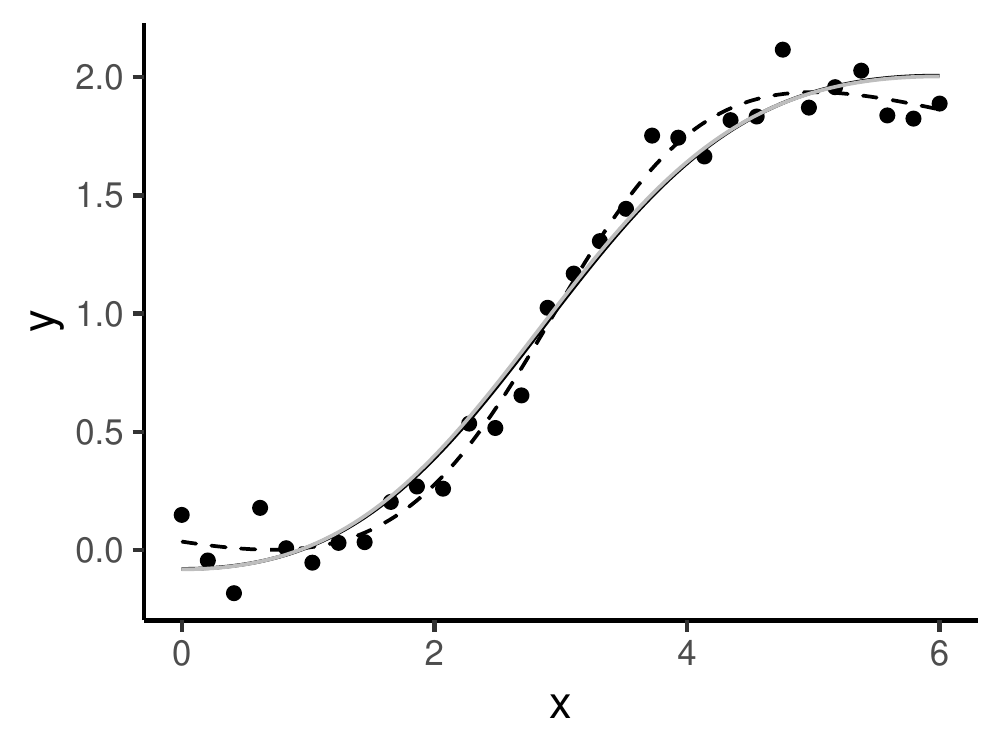}
    \caption{SMC-SA and CEPSO}
    \label{fig:ht0cepso}
  \end{subfigure}
  \begin{subfigure}[t]{0.49\textwidth}
    \centering
    \includegraphics[width=\textwidth]{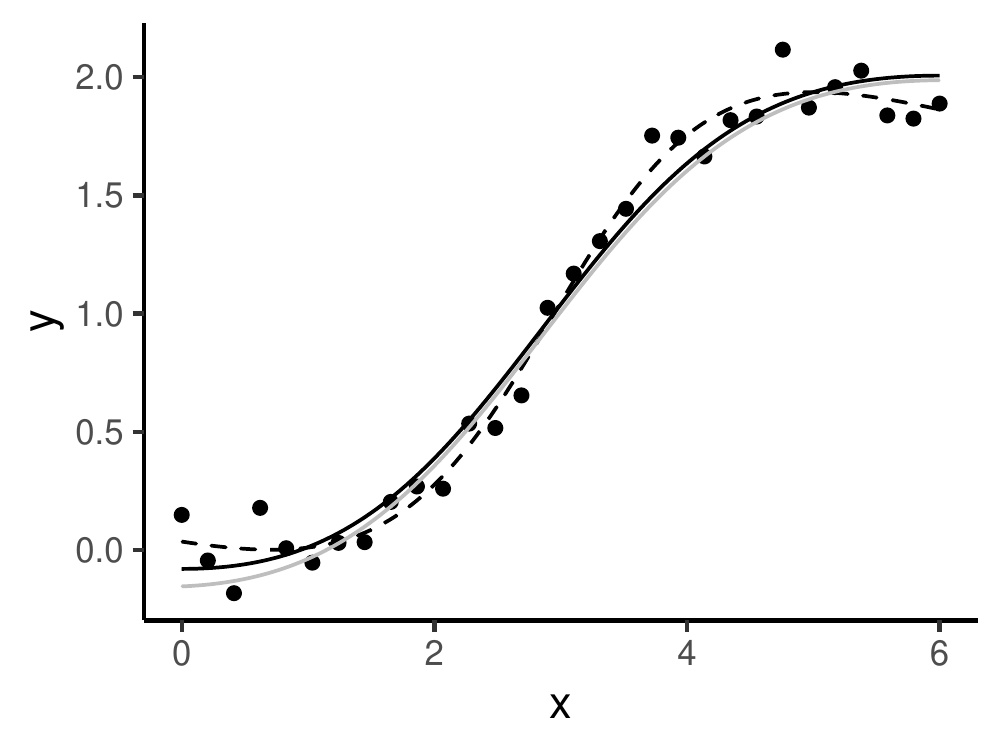}
    \caption{SMC-SA and multi-start SA}
    \label{fig:ht0multiSA}
  \end{subfigure}
  \caption{Comparisons between the best fit on HT0 from SMC-SA (black solid),
    multi-start SA (grey solid on the right) and CEPSO (grey solid on the left). The
    unconstrained least squares curve (black dotted), which is decreasing when $x > 5$,
    is shown as reference. The curve from CEPSO significantly overlaps that from
    SMC-SA.}
  \label{fig:ht0}
\end{figure}

We first consider a dataset generated from a hyperbolic tangent function which has a
sigmoidal shape
\begin{equation*}
  \text{HT0}: \quad y_i = 1 + \tanh{(x_i - 3)} + \epsilon_i \, ,
\end{equation*}
where $x_i = \frac{6}{29}(i - 1)$ and
$\epsilon_i \overset{i.i.d.}{\sim} \mathcal{N}(0, 0.3^2), i = 1, \ldots, 30$; see
Figure~\ref{fig:ht0}. This dataset will be referred to as HT0 for the remainder of this
paper. Our objective is to fit the following rational function model
\begin{equation}
  \label{eq:rational}
  r(x) = \dfrac{p_1(x)}{p_2(x)} = \dfrac{\beta_1 + \beta_2 x
    + \beta_3 x^2}{1 + \beta_4 x + \beta_5 x^2}, \quad y_i = r(x_i) + \epsilon_i, \quad
  \epsilon_i \sim \mathcal{N}(0, \sigma^2_\epsilon),
\end{equation}
which has a horizontal asymptote at $y = \frac{\beta_3}{\beta_5}$, to the HT0 dataset,
subject to a increasing monotonicity constraint in the interval $[0, 6]$. The necessary
conditions for this constraint are
\begin{equation}
  \label{eq:rational-mono}
  p'_1(x) p_2(x) - p_1(x) p'_2(x) \geq 0 , \quad \forall x \in [0, 6],
\end{equation}
and
\begin{equation}
  \label{eq:rational-con}
  p_2(x) \neq 0, \quad \forall x \in [0, 6],
\end{equation}
where \eqref{eq:rational-mono} is necessary for the first derivative of $r(x)$ to be
non-negative, and \eqref{eq:rational-con} for $r(x)$ to be continuous, in the interval
$[0, 6]$. Although both~\eqref{eq:rational-mono} and~\eqref{eq:rational-con} are not
closed-form expressions, these conditions can be verified
easily. For~\eqref{eq:rational-mono}, the left hand side of the inequality, which itself
is a polynomial, needs to fulfil two conditions: (i) all roots of the polynomial in
$[0, 6]$ must have even multiplicities; and (ii) the polynomial must evaluate to a
positive number at an arbitrary $x$ in $[0, 6]$. To verify~\eqref{eq:rational-con}, we
only need to ensure that the polynomial $p_2(x)$ does not have any roots in $[0, 6]$.

We use a least squares estimator, which minimises the model's residual sums of squares
\begin{equation}
  \label{eq:rational-rss}
  \ell(\bm{\theta}) = \sum_{i = 1}^{30}\left(y_i - r(x_i; \bm{\theta})\right)^2
  , \quad \bm{\theta} \in \mathcal{S},
\end{equation}
where $\bm{\theta} = (\beta_1, \beta_2, \beta_3, \beta_4, \beta_5)^T$, to obtain the
best fit of \eqref{eq:rational} to HT0. A rough estimate $\bm{\eta}_0$ for generating
starting states was obtained from a linear model rearranged from~\eqref{eq:rational}
\begin{equation}
  \label{eq:rational-rar}
  y = \beta_1 + \beta_2 x + \beta_3 x^2 - \beta_4 x y - \beta_5 x^2 y.
\end{equation}
The pilot estimate for CEPSO was an unconstrained least squares estimate obtained from
the Newton-Gauss algorithm implemented in \textsf{R}. The rough estimate $\bm{\eta}_0$
was used as the starting value for the Newton-Gauss algorithm.

\begin{table}
  \centering
  \caption{Mean, standard deviation, minimum, median and maximum of
    $\ell(\bm{\theta}^*)$ ; the number of chains with
    $\ell(\bm{\theta}^*) < 0.455 \times 1.01 = 0.460$; and the median running time in
    seconds, of 40 runs of the algorithms on HT0 (with a 2-point operator proposal for
    SA-based algorithms).}
  \begin{tabular}{*{1}{l}*{7}{c}}
    \firsthline
    Algorithms & Mean & SD & Min & Med & Max & \#Conv & Time \\
    \hline
    SMC-SA, reciprocal ($\alpha = 0.85$)
               & 1.224 & 0.61 & \textbf{0.455} & 1.649 & 1.781 & 14 & 326 \\
    SMC-SA, reciprocal ($\alpha = 0.95$)
               & 1.186 & 0.638 & \textbf{0.455} & 1.681 & 1.782 & 16 & 687 \\
    SMC-SA, logarithm
               & 1.565 & 0.236 & 0.464 & 1.614 & 1.801 & 0 & 636 \\
    multi-start SA
               & 1.185 & 0.313 & 0.532 & 1.329 & 1.585 & 0 & 293 \\
    CEPSO
               & 0.570 & 0.362 & 0.458 & 0.470 & 1.824 & 4 & 442 \\
    \lasthline
  \end{tabular}
  \label{tab:ht02c}
\end{table}

\begin{table}
  \centering
  \caption{Mean, standard deviation, minimum, median and maximum of
    $\ell(\bm{\theta}^*)$ ; the number of chains with $\ell(\bm{\theta}^*) < 0.460$;
    and the median running time in seconds, of 40 runs of the algorithms on HT0 with a
    1-point operator proposal.}
  \begin{tabular}{*{1}{l}*{7}{c}}
    \firsthline
    Algorithms & Mean & SD & Min & Med & Max & \#Conv & Time \\
    \hline
    SMC-SA, reciprocal ($\alpha = 0.85$)
               & 1.296 & 0.568 & \textbf{0.455} & 1.646 & 1.772 & 11 & 375 \\
    SMC-SA, reciprocal ($\alpha = 0.95$) 
               & 1.463 & 0.488 & \textbf{0.455} & 1.688 & 1.963 & 6 & 374 \\
    SMC-SA, logarithm
               & 1.615 & 0.113 & 1.363 & 1.639 & 1.767 & 0 & 296 \\
    multi-start SA
               & 1.296 & 0.201 & 0.509 & 1.328 & 1.558 & 0 & 309 \\
    \lasthline
  \end{tabular}
  \label{tab:ht01c}
\end{table}

For this example, we tried SMC-SA, multi-start SA and CEPSO with different
hyperparameters. The results from SA-based algorithms with a 2-point operator proposal
and CEPSO are presented in Table~\ref{tab:ht01c}. In terms of the best estimate
achieved, SMC-SA with a reciprocal schedule attained the lowest loss function value
($\ell(\bm{\theta}^*) = 0.455$), followed by CEPSO ($\ell(\bm{\theta}^*) =
0.458$). CEPSO generally has a higher chance of converging to the area near
$\bm{\theta}^*$ compared with SA-based algorithms, as shown in the medians of the loss
function values; although, SMC-SA with a reciprocal schedule can produce a more accurate
estimate, as demonstrated in the number of chains with a loss function value within 1\%
of the lowest $\ell(\bm{\theta}^*)$ achieved in this simulation study. The
hyperparameter $\alpha$ in the reciprocal schedule also did not have a significant
impact on the convergence of SMC-SA. However, SMC-SA with the logarithm schedule and
multi-start SA are less robust in the sense that the algorithms either did not converge
(i.e.~high median of~$\ell(\bm{\theta}^*))$ or produced inaccurate estimates (i.e.~no
chain getting close to the best $\bm{\theta}^*$). The results from SA-based algorithms
with a 1-point operator proposal are presented in Table~\ref{tab:ht01c}, which does not
indicate any significant difference in the summary statistics of loss function values
between using a 1-point and 2-point operator proposal. Nevertheless, SMC-SA with a
reciprocal schedule was more likely to produce accurate estimates when using a 2-point
operator proposal, as shown in the higher proportion of chains achieving
$\ell(\bm{\theta}^*) < 0.460$ (14 and 16 out of 40 in Table~\ref{tab:ht02c} compared to
11 and 6 out of 40 in Table~\ref{tab:ht01c}).

\begin{table}
  \centering
  \caption{Mean of $\ell(\bm{\theta}^*)$ of the 40 runs of the algorithms on HT0 (with a
    2-point operator proposal for SA-based algorithms) with different
    iterations. The $\alpha$ in the SMC-SA with a reciprocal schedule is set to
    0.95. The number of chains with $\ell(\bm{\theta}^*) < 0.460$ at a particular
    iteration is denoted in brackets.}
  \begin{tabular}{{l}*{6}{c}}
    \firsthline
    \multirow{2}{*}{Algorithms} & \multicolumn{6}{c}{Iterations}\\
    \cline{2-7}
    & 100 & 200 & 400 & 1000 & 2000 & 3000 \\
    \hline
    SMC-SA, recip. & 1.314 (0) & 1.231 (14) & 1.224 (14) & 1.224 (14)
                                             & NA & NA \\
    SMC-SA, log. & 1.683 (0) & 1.656 (0) & 1.634 (0) & 1.605 (0)
                                             & 1.565 (0) & 1.565 (0) \\
    multi-start SA & 1.764 (0) & 1.713 (0) & 1.639 (0) & 1.429 (0)
                                             & 1.241 (0) & 1.185 (0) \\
    CEPSO & 0.681 (0) & 0.598 (0) & 0.582 (1) & 0.574 (2)
                                             & 0.57 (4) & NA \\
    \lasthline
  \end{tabular}
  \label{tab:ht0records}
\end{table}

In Table~\ref{tab:ht0records}, we have also included the means of loss function values
and the number of chains achieving $\ell(\bm{\theta}^*) < 0.460$ when the algorithm
stopped at different iterations. The estimate generally improved slowly as the algorithm
iterates, with the exception of SMC-SA with a reciprocal schedule which achieved 14
estimates that were close to $\bm{\theta}^*$ in the first 400 iterations and plateaued
thereafter. This observation motivates using a reciprocal schedule instead of a
logarithm schedule to save computational resources, as we can terminate the SMC-SA well
before iterating 1000 times. However, we still iterated 1000 times in this simulation
study to ensure that all algorithms consume roughly the same amount of computational
resources.

The least squares curves produced by SMC-SA, multi-start SA and CEPSO are illustrated in
Figure~\ref{fig:ht0}, with the unconstrained least squares curve (black dotted) serving
as a reference. Since the results produced by SMC-SA and CEPSO did not differ greatly,
their regression curves are overlaying one another, and thus appear to be a single
curve. Conversely, there is a noticeable difference between the curves produced from
SMC-SA and multi-start SA, as the best estimate attained by the latter was
sub-optimal. From these results, we conclude that SMC-SA can produce estimates
comparable to that from CEPSO but without the need of a pilot estimate.

\subsubsection{Modelling the growth of Cucumis melo}
Rational function models with the same degree of polynomial on the numerator and
denominator have a horizontal asymptote, which can be useful when modelling growth
curve. For instance, \eqref{eq:rational} will converge to $y = \frac{\beta_3}{\beta_5}$
when $x \rightarrow \infty$. In this example, we demonstrate the practical usage of
rational function models by using \eqref{eq:rational} to model the growth of
\textit{Cucumis melo} seeds that were grown at $15^{\circ}$C. The dataset comprises 15
mean observations of the height of \textit{Cucumis melo} seeds recorded over 24 days;
more details can be found in \citet{pearl34growth}. The response (\texttt{height}) and
covariate (\texttt{day}) are scaled appropriately for numerical stability.

We constrained the growth curve to be monotonically increasing between the first and
last recorded dates as we do not expect the seedlings to reduce in height over time. A
least squares estimator was employed to fit \eqref{eq:rational}, and $\bm{\eta}_0$ and a
pilot estimate were obtained using the same procedure in
Section~\ref{sec:hyper-tangent}.

\begin{table}
  \centering
  \caption{Mean, standard deviation, minimum, median and maximum of
    $\ell(\bm{\theta}^*)$ ; the number of chains with
    $\ell(\bm{\theta}^*) < 0.246 \times 1.02 = 0.251$; and the median running time in
    seconds, of 40 runs of the algorithms on \textit{Cucumis melo} (with a 2-point
    operator proposal for SA-based algorithms).}
  \begin{tabular}{*{1}{l}*{7}{c}}
    \firsthline
    Algorithms & Mean & SD & Min & Med & Max & \#Conv & Time \\
    \hline
    SMC-SA, reciprocal ($\alpha = 0.85$)
               & 0.43 & 0.227 & \textbf{0.246} & 0.341 & 1.112 & 10 & 387 \\
    SMC-SA, reciprocal ($\alpha = 0.95$)
               & 0.441 & 0.259 & \textbf{0.246} & 0.382 & 1.451 & 7 & 386 \\
    SMC-SA, logarithm
               & 0.639 & 0.626 & 0.247 & 0.373 & 3.029 & 6 & 397 \\
    CEPSO
               & 0.295 & 0.017 & 0.274 & 0.289 & 0.337 & 0 & 407 \\
    \lasthline
  \end{tabular}
  \label{tab:melon}
\end{table}

\begin{figure}
  \centering
  \begin{subfigure}[t]{0.49\textwidth}
    \centering
    \includegraphics[width=\textwidth]{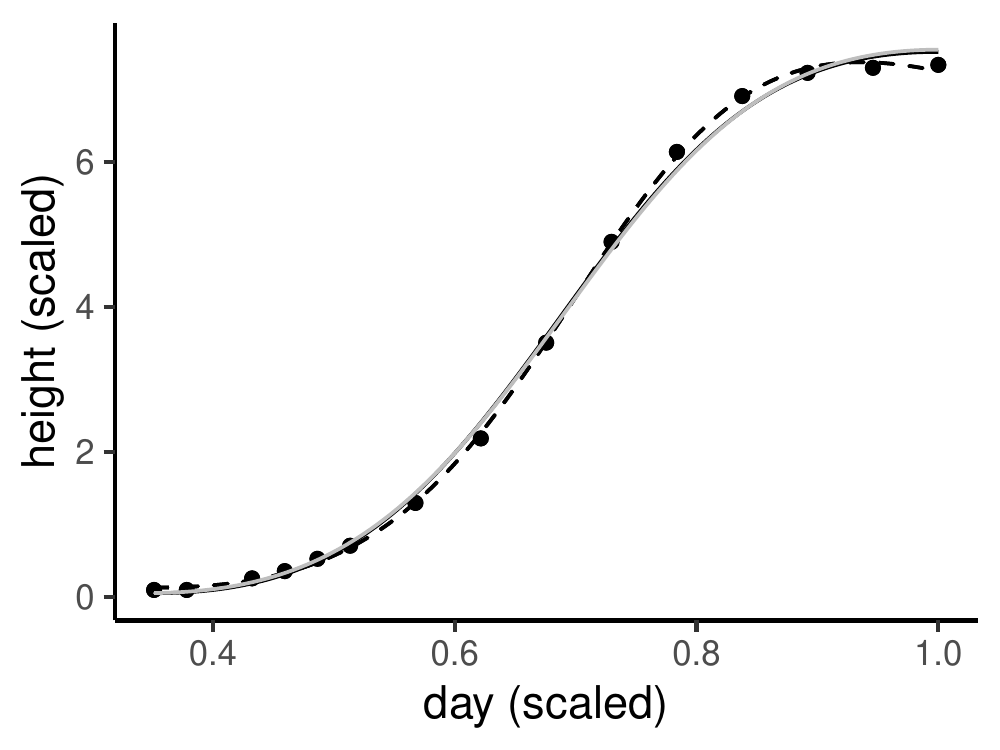}
    \caption{SMC-SA and CEPSO}
    \label{fig:meloncepso}
  \end{subfigure}
  \begin{subfigure}[t]{0.49\textwidth}
    \centering
    \includegraphics[width=\textwidth]{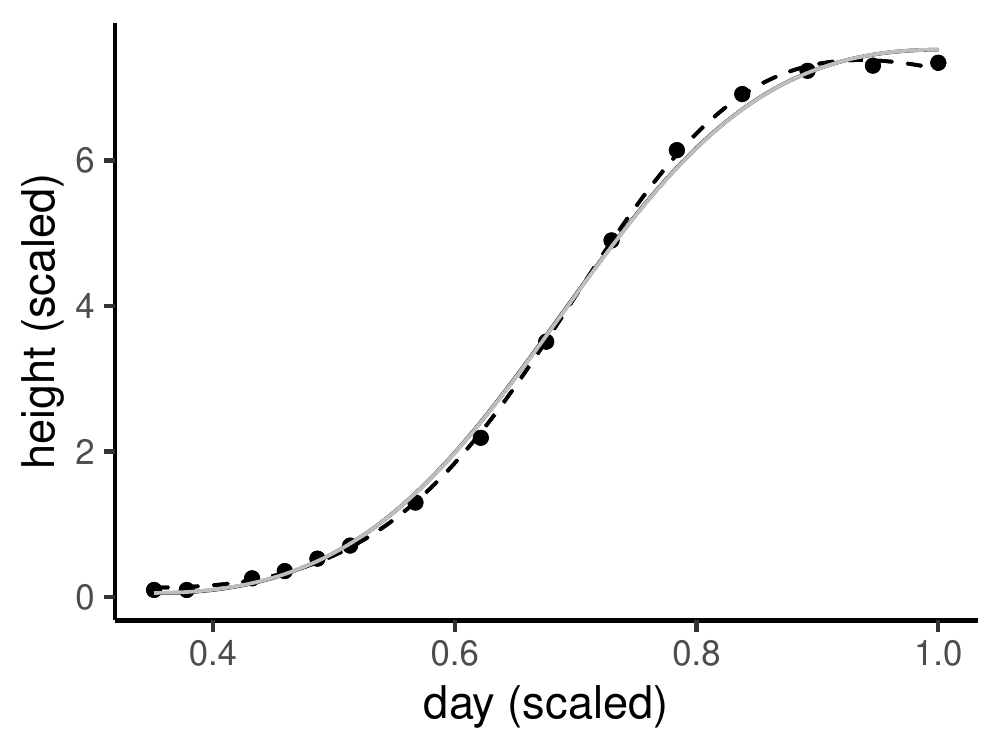}
    \caption{SMC-SA with different schedules}
    \label{fig:melonmultiSA}
  \end{subfigure}
  \caption{Comparisons between the growth curves of \textit{Cucumis melo} seeds. The
    panel on the left shows the best curves from SMC-SA with a reciprocal schedule
    (black solid) and CEPSO (grey solid), while the right compares SMC-SA with
    reciprocal (black solid) and logarithm (grey solid) schedules. The unconstrained
    least squares curve (black dotted) is shown as reference. The grey curves from CEPSO
    (left panel) and SMC-SA with the logarithm schedule (right panel) significantly
    overlap the black curves produced from SMC-SA with a reciprocal schedule; hence, the
    plots appear as if there are only grey curves.}
  \label{fig:melon}
\end{figure}

The results from SMC-SA and CEPSO are presented in Table~\ref{tab:melon}. Contrary to
the previous example in Section~\ref{sec:hyper-tangent}, the cooling schedule in SMC-SA
did not affect the convergence noticeably. SMC-SA did manage to produce estimates with a
lower loss function value compared with CEPSO, which failed to produced an estimate with
a loss function value close to 0.246, which is the lowest loss function value achieved
in this simulation. The estimates produced by CEPSO, nonetheless, were more stable
compared with that from SMC-SA, in the sense that the estimates are similar to each
other. The best regression curves produced by each algorithm are also shown in
Figure~\ref{fig:melon}, which does not indicate any visually noticeable difference
between the curves produced.

\subsection{Constrained robust models}
\label{sec:robust-models}

Observed data are sometimes contaminated with outliers, and a different loss
function can be used to reduce their influence on the model fit. In this example, we
demonstrate that our algorithm is capable of estimating the regression coefficients of
\eqref{eq:rational}, subject to both \eqref{eq:rational-mono} and
\eqref{eq:rational-con}, with an M-estimator \citep{huber81robust} which minimises the
following loss function
\begin{equation}
  \label{eq:rational-mes}
  \ell(\bm{\theta}) = \sum_{i = 1}^{30} \rho \left(y_i - r(x_i; \bm{\theta})\right),
  \quad \bm{\theta} \in \mathcal{S},
\end{equation}
where $\rho(\cdot)$ is carefully chosen to limit the influence of outliers on the
overall model fit. The least squares estimator is essentially a special case of
\eqref{eq:rational-mes} where $\rho(u) = u^2$, but this particular choice of
$\rho(\cdot)$ greatly inflates the influence of outliers. Here, we employ Tukey's
biweight loss function
\begin{equation*}
  \rho(u) = 
  \begin{dcases}
    \dfrac{c^2}{6}\left( 1-\left( 1-\left( \dfrac{u}{c} \right)^2 \right)^3 \right), &
    \lvert u \rvert \leq c \, , \\
    \dfrac{c^2}{6}, & \lvert u \rvert > c \, , \\
  \end{dcases}
\end{equation*}
where the choice of $c$ is arbitrary, but our default value is 1.

\begin{figure}
  \centering
  \includegraphics[width=.7\textwidth]{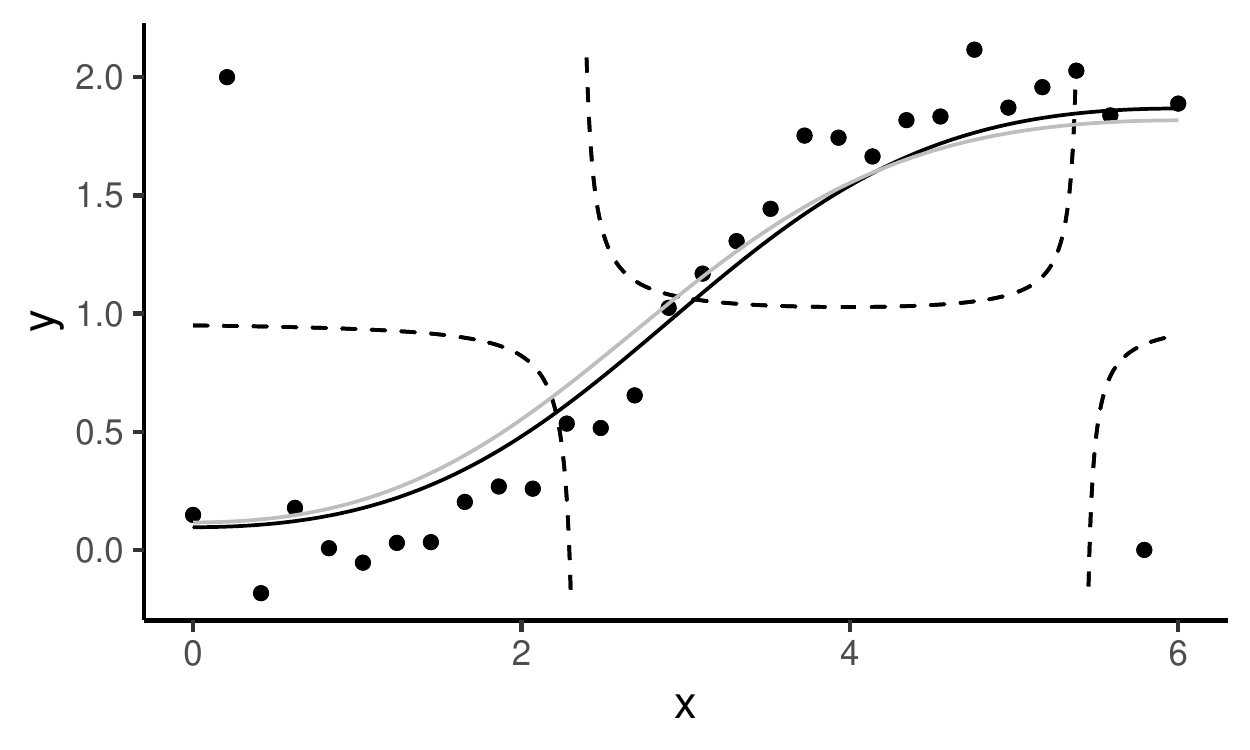}
  \caption{A comparison between the best fit on HT1 from SMC-SA (black solid) and CEPSO
    (grey solid). There is a noticeable difference between the two curves. The
    unconstrained least squares curve (black dotted), which has poles at $x = 2.35$ and
    $x = 5.42$, is shown as reference.}
  \label{fig:ht1}
\end{figure}

In this example, we used a dataset purposefully contaminated with outliers. This dataset
will be referred to as HT1, and was produced by setting the values of $y_2$ and $y_{28}$
in HT0 to 2 and 0 respectively. We used an unconstrained least squares estimate of
\eqref{eq:rational} fitted on HT0 as the pilot estimate of CEPSO. The estimate was
obtained using the Newton-Gauss algorithm starting from $\bm{\eta}_0$, which was
obtained following the same procedure described in
Section~\ref{sec:hyper-tangent}. However, due to the presence of outliers, the estimate
obtained from the Newton-Gauss algorithm (the black dotted curve in
Figure~\ref{fig:ht1}) is a local minimum in the corresponding residual sum of squares
function, and thus may not be suitable to be used as the pilot estimate. An
unconstrained M-estimate can also be used, but to obtain this will require a different
optimisation method.

\begin{table}
  \centering
  \caption{Mean, standard deviation, minimum, median and maximum of
    $\ell(\bm{\theta}^*)$ ; the number of chains with
    $\ell(\bm{\theta}^*) < 3.439 \times 1.01 = 3.473$; and the median running time in
    seconds, of 40 runs of the algorithms on HT1 with a 2-point operator proposal.}
  \begin{tabular}{*{1}{l}*{7}{c}}
    \firsthline
    Algorithms & Mean & SD & Min & Med & Max & \#Conv & Time \\
    \hline
    SMC-SA, reciprocal ($\alpha = 0.85$)
               & 3.875 & 0.466 & \textbf{3.439} & 3.441 & 4.426 & 21 & 365 \\
    SMC-SA, reciprocal ($\alpha = 0.95$)
               & 3.958 & 0.455 & \textbf{3.439} & 4.188 & 4.429 & 16 & 680 \\
    SMC-SA, logarithm
               & 4.420 & 0.032 & 4.289 & 4.425 & 4.454 & 0 & 255 \\
    CEPSO
               & 3.573 & 0.149 & 3.489 & 3.567 & 4.452 & 0 & 359 \\
    \lasthline
  \end{tabular}
  \label{tab:ht1}
\end{table}

In this example, SMC-SA with a reciprocal schedule outperformed CEPSO in terms of the
minimum loss function value achieved in 40 runs, as demonstrated in
Table~\ref{tab:ht1}. There is also a noticeable difference in the regression curves
produced by SMC-SA and CEPSO (Figure~\ref{fig:ht1}). Since we were using a sub-optimal
pilot estimate, we expect that CEPSO would struggle to produce accurate estimates when
comparing with SMC-SA. Therefore, in the absence of a satisfactory pilot estimate, we
conclude that SMC-SA is preferable.

\subsection{Constrained B-spline models}
\label{sec:bs-models}
SMC-SA is not restricted to fitting rational function models. In this section, we
demonstrate our algorithm by fitting a quadratic B-spline model on a dataset collected
from a light detection and ranging (LIDAR) experiment \citep{sigrist94air}. The data
comprises 221 observations of log-ratios of received light from two laser sources
(\texttt{logratio}), recorded against the distance travelled before the light is
reflected back to its source (\texttt{range}). The log-ratio is expected to decrease as
the range increases. For the purpose of this example, both response and covariate were
divided by their respective maximum to ensure numerical stability.

\begin{figure}
  \centering
  \includegraphics[width=\textwidth]{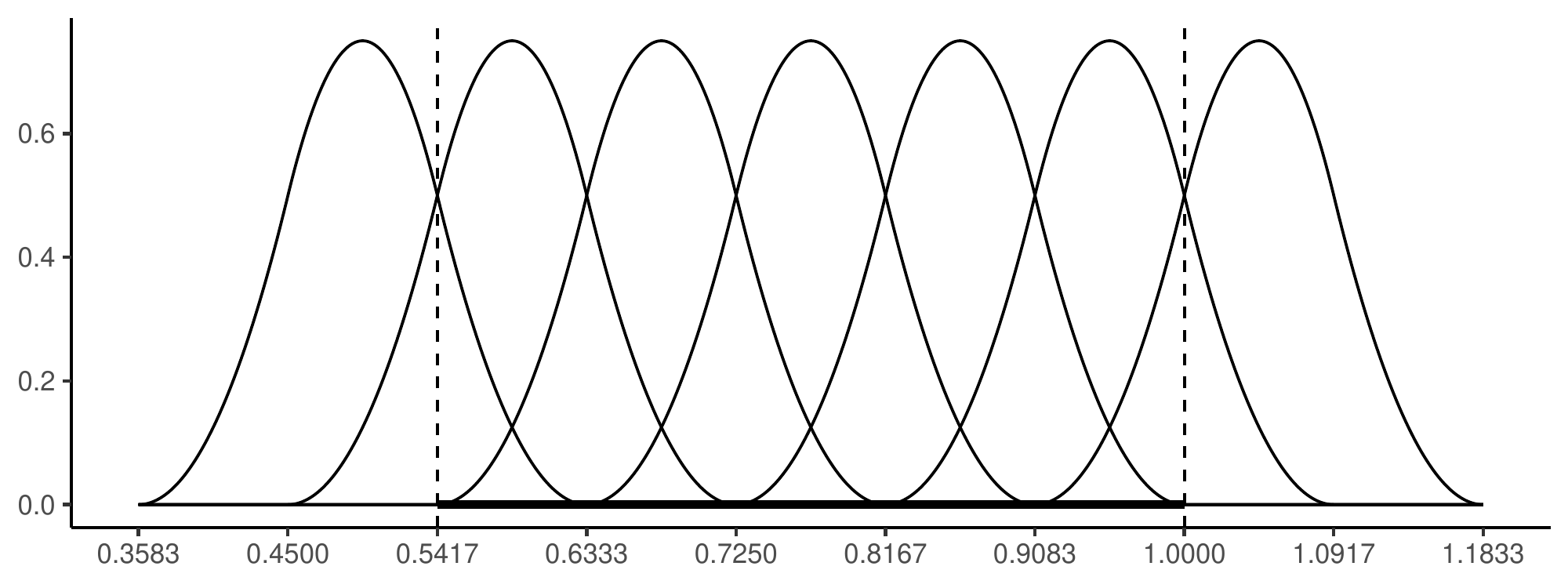}
  \caption{Quadratic B-spline basis functions with equidistant knots. The knots
    positions are marked on the horizontal axis.}
  \label{fig:basis}
\end{figure}

The mathematical formulation of the quadratic B-spline model is given by
\begin{equation}
  \label{eq:lidar}
  \texttt{logratio}_i = \sum^{7}_{j = 1} \beta_j B_j(\texttt{range}_i) + \epsilon_i,
  \quad \epsilon_i \sim \mathcal{N}(0, \sigma^2_{\epsilon}), \quad i = 1, \ldots, 221,
\end{equation}
where $B_j(\cdot)$ denotes quadratic B-spline basis functions with 10 equidistant knots
spanning between the minimum ($\texttt{range}_{min} = 0.5417$) and maximum
($\texttt{range}_{max} = 1$) of the covariate; see Figure~\ref{fig:basis}. For the
quadratic B-spline model to be monotonically decreasing over the interval
\mbox{$\texttt{range} \in [0.5417, 1]$}, we require the regression coefficients to
follow $\beta_7 \geq \beta_6 \geq \ldots \geq \beta_1$; further details are given
in~\ref{sec:monotonicity}. We used a least squares estimator to fit \eqref{eq:lidar} and
chose an arbitrary sequence $\bm{\eta}_0 = (7, 6, \ldots, 1)^T$ as the rough estimate
for generating starting states. The ordinary least squares estimate of \eqref{eq:lidar}
was employed as the pilot estimate of CEPSO and admits a close-form expression
$\widehat{\bm{\beta}}_{OLS} = (\bm{B}^T\bm{B})^{-1}\bm{B}^T\bm{y}$ where
$\bm{B} = (\bm{B}_1 \cdots \bm{B}_7)$,
$\bm{B}_j = (B_j(\texttt{range}_1), \cdots, B_j(\texttt{range}_{221}))^T$,
$j = 1, \ldots, 7$, and the response vector
$\bm{y} = (\texttt{logratio}_1, \cdots, \texttt{logratio}_{221})^T$.

\begin{table}
  \centering
  \caption{Mean, standard deviation, minimum, median and maximum of
    $\ell(\bm{\theta}^*)$ ; the number of chains with
    $\ell(\bm{\theta}^*) < 1.530 \times 1.01 = 1.545$; and the median running time in
    seconds, of 40 runs of the algorithms on LIDAR with a 2-point operator proposal.}
  \begin{tabular}{*{1}{l}*{7}{c}}
    \firsthline
    Algorithms & Mean & SD & Min & Med & Max & \#Conv & Time \\
    \hline
    SMC-SA, reciprocal ($\alpha = 0.85$)
               & 1.530 & $<$ 0.001 & \textbf{1.530} & 1.530 & 1.530 & 40 & 752 \\
    SMC-SA, reciprocal ($\alpha = 0.95$)
               & 1.530 & $<$ 0.001 & \textbf{1.530} & 1.530 & 1.530 & 40 & 762 \\
    SMC-SA, logarithm
               & 1.540 & 0.003 & 1.533 & 1.54 & 1.547 & 37 & 2235 \\
    CEPSO
               & 1.530 & $<$ 0.001 & \textbf{1.530} & 1.530 & 1.530 & 40 & 682 \\
    \lasthline
  \end{tabular}
  \label{tab:lidar}
\end{table}

\begin{figure}
  \centering
  \includegraphics[width=0.7\textwidth]{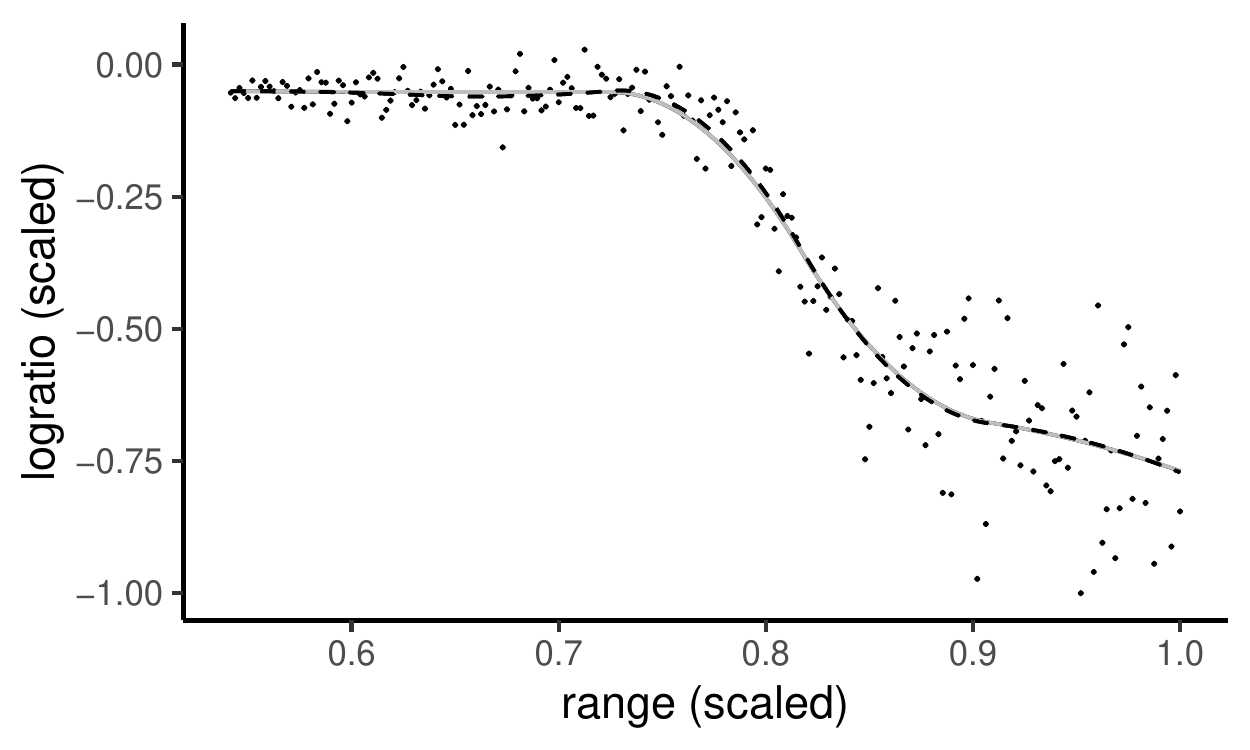}
  \caption{Fitting a quadratic B-spline model on the LIDAR dataset. The curve from CEPSO
    (grey solid) completely overlaps that from SMC-SA with a reciprocal schedule (black
    solid, not visible). The unconstrained least squares curve (black dotted) is very
    similar to the constrained estimate; however, the unconstrained curve has an
    increasing trend roughly in $\texttt{range} \in [0.65, 0.75]$. }
  \label{fig:lidar}
\end{figure}

In this example, SMC-SA with a reciprocal schedule produced indistinguishable results to
CEPSO, as shown in Table~\ref{tab:lidar}, although CEPSO is slightly faster in terms of
computation time. More remarkably, all 40 runs of these algorithms converged to the same
estimate. This accomplishment is partly attributed to the similarity between the
unconstrained and constrained estimates, which can be seen in
Figure~\ref{fig:lidar}. However, SMC-SA with a logarithm schedule performed slightly
worse than the rest in terms of the quality of estimates and took significantly more
time to run. This inefficiency is mainly due to the relatively high variance of the
proposal of SMC-SA with the logarithm schedule, resulting in the algorithm spending too
much effort on exploring rather than refining the estimate. From this result, we can
conclude that SMC-SA with a reciprocal schedule is a competitive alternative to CEPSO,
with the benefit of not requiring a pilot estimate to operate.

\section{Discussion}
\label{sec:discussion}
We introduced a constrained-augmented SMC-SA algorithm which is capable of optimising
with a wide variety of constraints, including those that do not have closed-form
expressions. Our algorithm only requires an indicator function that verifies the
feasibility of an estimate to operate, as opposed to CEPSO which requires an additional
pilot estimate which may be difficult to obtain.

Our algorithm was built upon the fact that a Boltzmann distribution will converge to a
Dirac measure at the global minimum as the cooling temperature decreases to 0, even when
its support is restricted to a subset of a real space. Therefore, we followed the idea
of \citet{zhou13sequential} to sample from the sequence of Boltzmann distributions using
a SMC sampler. We could similarly use a sampler that incorporates Hamiltonian dynamics,
which may explore high-dimensional parameter space more efficiently, or parallel
tempering to make our algorithm more robust against local minima. Furthermore, our
algorithm can potentially be sped up by augmenting the Boltzmann distribution, as in
\citet{liang14simulated}.

The running time of our algorithm is mainly concentrated on the rejection sampler that
generates proposed states from \eqref{eq:proposal-cons}. However, the usage of rejection
samplers, which is necessary in our algorithm due to the lack of assumptions on the
nature of the constraints, is notoriously inefficient for truncating high dimensional
distributions, particularly when the probability mass of the untruncated
\eqref{eq:proposal-cons} is spreading mainly across the infeasible set. For this reason,
it is advisable to start the algorithm with feasible starting values to avoid lengthy
computation at the first iteration, although our algorithm works with arbitrary starting
values in principle. Moreover, when working with a large set of starting values, the
computational time can be further improved by parallelising the SA-move, which is
executed independently for each $\bm{\gamma}^j_k$.

Although our algorithm is originally intended to solve shape-constrained regression
problems, it can be generalised to solve generic constrained optimisation
problems. However, one may expect that a reciprocal schedule, which is suggested for
shape-constrained regression, may be less suitable since the cooling schedule is usually
dependent on the problem at hand. In such cases, we advise a slight modification to the
$\alpha$ in \eqref{eq:recip-schedule} before trying an alternative family of cooling
schedules when a performance improvement is not immediately attained
\citep{fouskakis02stochastic}.

The proposal variance, in particular its decreasing factor, is also crucial to the
algorithm convergence. Our motivation of decreasing the variance after each iteration is
to maintain a healthy acceptance rate (around 0.2 to 0.5) at the latter stages of the
algorithm, when a substantial move from the current states is unlikely to improve the
search chain \citep{zhou13sequential}. During that period, the algorithm should focus on
refining the current estimates rather than exploring new regions in the parameter space,
and a small proposal variance will facilitate this effort. However, the choice of a
$3\%$ (or $0.2\%$ when using the logarithm schedule) decrement of the
proposal~$\sigma^2$ in our study may be too aggressive in some situations and prevent
the algorithm to explore the parameter space effectively. Hence, the user should adjust
the decrement accordingly.

As with all optimisation problems, rescaling of the data is sensible to ensure numerical
stability which is reflected in the choice of the scale of our example data presented,
which in turn affects the choice of hyperparameters (i.e.~cooling schedule and proposal
variance).

\newpage
\bibliographystyle{elsarticle-harv}
\bibliography{./mainbib}

\newpage
\appendix
\section{Monotone B-spline models}
\label{sec:monotonicity}
\begin{figure}
  \centering
  \includegraphics[width=\textwidth]{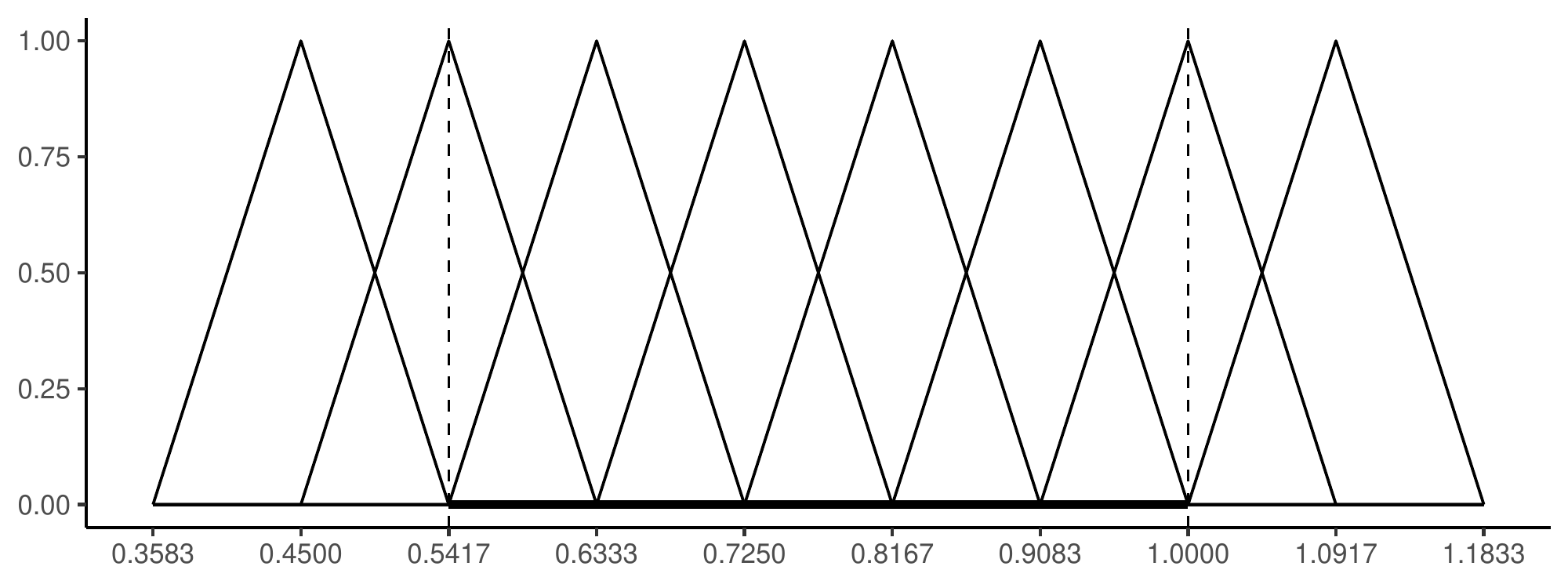}
  \caption{The linear B-spline basis functions that corresponds to the first order
    derivative of the quadratic basis functions in Figure~\ref{fig:basis}. The locations
    of the knots are marked on the horizontal axis.}
  \label{fig:deriv}
\end{figure}

A B-spline model with degree $d$ and a set of knots
\mbox{$k_1 < k_2 < \ldots < k_{J+d+1}$} can be written as
\begin{equation*}
  f(x) = \sum^J_{j = 1} \beta_j B^d_j(x),
\end{equation*}
where $f$ only evaluates in $x \in [k_{d+1}, k_{J+1}]$. Suppose that knots are
equidistant and separated by a distance $h$, the basis functions can be derived from the
following recursive formula \citep{deboor78practical}
\begin{align*}
  B^0_j(x) &= \mathds{1}_{x \in [k_{j}, k_{j+1}]} \quad &\text{if $d = 0$}; \\
  B^d_j(x) &= \dfrac{x - k_{j}}{hd} B^{d-1}_j(x) + \dfrac{k_{j+ d + 1} - x}{hd}
             B^{d-1}_{j+1}(x) \quad &\text{if $d \geq 1$}.
\end{align*}
The corresponding first order derivative of $B^d_j(x)$ is given by
\begin{equation*}
  \dv{x} B^d_j(x) = \dfrac{1}{h} \left(B^{d-1}_j(x) - B^{d-1}_{j+1}(x)\right),
\end{equation*}
and consequently, $f'(x)$ can be deduced
\begin{align}
  f'(x) &= \dfrac{1}{h} \sum^J_{j = 1} \beta_j \left(B^{d-1}_j(x) -
          B^{d-1}_{j+1}(x)\right) \nonumber \\
        &= \dfrac{1}{h} \left( \beta_1 B^{d-1}_1(x) - \beta_J B^{d-1}_{J+1}(x) +
          \sum^J_{j = 2} (\beta_j - \beta_{j-1}) B^{d-1}_j(x) \right) \nonumber \\
        &= \dfrac{1}{h} \sum^J_{j = 2} (\beta_j - \beta_{j-1}) B^{d-1}_j(x) , \quad
          \text{if $x \in (k_{d+1}, k_{J+1})$}. \label{eq:1st-deriv-bs}
\end{align}
The third equality follows from the fact that we are only interested in evaluating $f$
for $x \in [k_{d+1}, k_{J+1}]$, and that $B^{d-1}_1(x)$ and $B^{d-1}_{J+1}(x)$ evaluate
to 0 over that interval. For quadratic B-spline models ($d = 2$), we deduce that
necessary conditions for decreasing monotonicity,
$f'(x) \leq 0, \forall x \in (k_{d+1}, k_{J+1})$, can be achieved by ensuring that the
coefficients of the linear B-spline basis functions in \eqref{eq:1st-deriv-bs},
$\beta_j - \beta_{j-1}$, are non-positive since B-spline basis functions are
non-negative everywhere; otherwise, $f'(x)$ will evaluate to a positive value at at
least one knot $x \in \{k_{d}, \ldots, k_{J} \}$; see Figure~\ref{fig:deriv}. Therefore,
the necessary condition for a monotonically decreasing quadratic spline is
$\beta_1 \geq \beta_2 \geq \ldots \geq \beta_{J}$. The condition for increasing
monotonicity can also be deduced similarly.

\end{document}